\documentclass[12pt,BCOR=0mm,DIV=35]{article}
\usepackage{latexsym}
\usepackage{graphicx}
\usepackage{amsmath}
\usepackage{amsfonts}
\usepackage{amssymb}
\numberwithin{equation}{section}
 
\newcommand{\mR}{\mathbb R}
\newcommand{\be}[1]{\begin{equation}\label{e#1}}
\newcommand{\ee}{\end{equation}}

\title{Quantum conformal mechanics}
\author{Krzysztof  Andrzejewski
\\ \\
\small Department of  Computer Science, \\
\small University of \L\'od\'z,\\
\small Pomorska 149/153, 90-236 {\L}\'od\'z, Poland\\
\small E-mail: k-andrzejewski@uni.lodz.pl
}
\date{}
\begin{document}
\maketitle
\begin{abstract}
The quantum mechanics of one degree of freedom exhibiting  the exact conformal $SL(2,\mR)$ symmetry is presented. The starting point is the classification  of the  unitary irreducible  representations of the $SL(2,\mR)$ group (or, to some extent, its universal covering). The coordinate representation is defined as the basis  diagonalizing  the special conformal generator  $\hat K$. It is indicated how the resulting theory emerges  from the canonical/geometric quantization of the Hamiltonian dynamics  on the relevant  coadjoint orbits.
\end{abstract}
\section{Introduction}
The $SL(2,\mR)$ group provides the prototype of conformal  groups.  It describes conformal transformations of $(1+0)$-dimensional space-time; the generators  of $SL(2,\mR)$ corresponds to time translations, dilatations and special conformal transformations. The dynamical model  exhibiting such  a symmetry, the so called  conformal mechanics, has been introduced  in Refs. \cite{b1} and \cite{b2} (for their supersymmetric extensions see \cite{b3,b4}). Since then there appeared  number of papers  devoted to the detailed study of the  conformal mechanics \cite{b3}-\cite{b37b}, both on classical and quantum levels. 
\par
The basic quantities entering the description of conformal mechanics are the Hamiltonian $H$, the dilatation generator $D$ and the generator of conformal transformations $K$: 
\begin{align}
\label{e1}
H&=\frac{p^2}{2m}+\frac{g^2}{2x^2},\\
\label{e1a}
D&=-\frac{1}{2}xp,\\
\label{e1b}
K&=\frac{m}{2}x^2,
\end{align}
where $x$  and $p$ are standard canonical variables, $m$ -- the mass of  a particle (which we will put  equal to one in what follows) and  the coupling constant  $g$. They obey $SL(2,\mR)$ commutations rules with respect to the standard Poisson brackets
\be{2}
\{D,H\}=-H,\quad \{D,K\}=K,\quad \{K,H\}=-2D.
\ee
This structure survives on the quantum level provided $D$ is ordered appropriately. The value of  the the Casimir operator can be expressed in terms of the  coupling constant as follows
\be{3}
C=HK-D^2=\frac{g^2}{4},
\ee 
or on the quantum level ($\hbar=1$),
\be{4}
\hat C=\frac{1}{2}(\hat H\hat K+\hat K\hat H)-\hat D^2=\frac{g^2}{4}-\frac{3}{16}.
\ee
\par 
The main restriction usually imposed on the Hamiltonian  $H$ is that the potential is repulsive, $g^2>0$. Then the phase space $P$  is a half-plane $\{(x,p) |x\in\mR_+,p\in \mR\}$. This phase space is complete in the sense that, given any initial point, $(x(0),p(0))\in P$, the whole trajectory $(x(t),p(t))\in P$, $-\infty<t<\infty$, belongs to $P$. This is no longer the case for the non-positive  coupling constant. If $g=0$ we have the free motion so the natural candidate  for the phase space is the whole plane. The situation gets even more complicated  when $g$  becomes negative. The singularity of the potential  at $x=0$ defines the boundary of the phase space which again becomes the half-plane $x>0$, $p\in \mR$. However, it is now no longer complete due to the "falling on the center" phenomena: the particle can  reach the boundary of phase space at finite time.
\par
This trouble has its counterpart on the quantum level. Upon quantization the Hamiltonian becomes the differential operator of the second order. Actually, it is  a formal differential expression which becomes  meaningful provided it defines a self-adjoint operator acting in the Hilbert space of states. To this end one  has to impose the appropriate boundary conditions yielding such an operator \cite{b38}-\cite{b41}. It appears that in many cases  (including the one considered here) such a procedure is not unique. In fact, for the conformal Hamiltonian (\ref{e1}) one  arrives at the following classification \cite{b42}: 
\begin{enumerate}
\item[-] for $g^2\geq\frac34$ there exists only one self-adjoint  Hamiltonian (\ref{e1}); the energy spectrum is purely continuous  and extends over the whole nonnegative semiaxis $E\geq0$;
\item[-] for $\frac{3}{4}>g^2>-\frac14$ there exists a one-parameter family of self-adjoint  Hamiltonians; the physical interpretation of this fact  in terms of regularized  cut-off potential  is discussed in the  classical  Landau and Lifschitz  book \cite{b43}; however, such an interpretation does not cover all possibilities \cite{b42}.  In the case under consideration the continuous part of the energy spectrum again  extends over the whole nonnegative semiaxis. However, for certain self-adjoint extensions there exists an additional single bound state  of nonnegative energy (surprisingly enough, this  can happen even in the region  $\frac{3}{4}>g^2\geq 0 $, i.e., for the repulsive potential); 
\item[-] for $g^2=-\frac14$  again there exists  a one-parameter family of self-adjoint Hamiltonians. The spectrum  consists  of continuous  part $E\geq 0$  together  with a single bound state of negative energy.
\item[-] for $g^2<-\frac 14$ there exists also a one parameter  family. The continuous spectrum  consists of  all nonnegative   energies; there exists an infinite family of negative   energy bound states  concentrating  exponentially to zero and going exponentially to $-\infty$.
\end{enumerate}
In all cases, except the first one, the self-adjoint Hamiltonian is defined by dimensionful parameter which sets the scale of the bound state(s). Therefore, the scale symmetry must be broken as well as the conformal one. Let us note that it is not a priori  clear whether their generators can be defined as genuine self-adjoint operators  obeying $sl(2,\mR)$ algebra except in formal sense  and whether the resulting algebra can be integrated to yield the representation of the (covering)  of the  $SL(2,\mR)$   group. The question arises whether the exists a quantum mechanical model carrying genuine (i.e., unbroken) $SL(2,\mR)$ symmetry.  The answer seems to be positive for strong enough repelling  potentials. In this case both on the classical and quantum level   we are dealing with conformal invariant systems. However, in other cases the situation becomes  more complicated; classically, the accessible phase  space becomes incomplete in the sense described above while quantum mechanically trouble arises with scale invariant definitions of self-adjoint group generators.
\par
In the present paper we construct fully conformally invariant quantum mechanics for all values of the coupling constant. The starting point is the construction of conformal mechanics  as described in Refs. \cite{b33}  and \cite{b37} where  the general and elegant method of coadjoint orbits,  which allows to define Hamiltonian dynamics invariant under a transitive action of a given symmetry group, has been applied.  The main  conclusion  following from the construction presented there  is that, at least on the classical level, the troubles arising in the case of nonrepelling potential are the artifacts arising due to the nontrivial topology of the phase space.  Once this is properly recognized  the motion becomes completely regular  and the  singularity at $x=0$ appears to be completely  spurious as resulting from the fact that the phase space cannot be  covered by  a single map. So, on the classical level we are dealing  with perfectly regular (in fact, rather simple) dynamical system.  The only inconvenience is that, in general, there exists no globally defined system of Darboux coordinates.
\par         
Our aim in the present paper is to quantize the classical dynamics defined   with  the help of the orbits method. We show that for all values of the coupling constant one can find the relevant quantum mechanical  system exhibiting exact $SL(2,\mR)$ conformal symmetry. Its Hilbert space of states spans an irreducible unitary representation of the $SL(2,\mR)$ group (or its universal covering). As it is known \cite{b44}-\cite{b46} the irreducible unitary representations  of $SL(2,\mR)$  can be  classified as follows. First, there exists  the continuous serie characterized  by the pairs  $(\rho,\epsilon)$, where $\epsilon=0,1$; $\rho\in\mR$  for $\epsilon=0$ while $\rho\in\mR\backslash\{0\}$   for $\epsilon=1$; two such representations $(\rho,\epsilon)$, $(\rho',\epsilon')$ are equivalent  if $\epsilon=\epsilon'$ and $\rho=\pm\rho'$.  Second, there are discrete series characterized by the integers $m\geq 1$; they are all inequivalent. Further, there is  the supplementary  serie   indexed by $\rho\in(-1,1)$, $\rho\neq 0$; again  the representations corresponding to $\rho$  and $-\rho$  are unitary  equivalent. Finally, there are two mock representations which can be viewed as the formal limits $m\rightarrow 0^+$ of the discrete ones  or as two irreducible components of the representation $(0,1)$  of the continuous serie.
\par
 All these representations are used as the building blocks in our construction. Due to the equation (\ref{e4}) one can relate the values of coupling constants and Casimir operators  which allows to identify the representations arising for particular values of coupling. As a result  the following picture emerges:
\begin{center}
\setlength{\unitlength}{1mm}
\begin{picture}(130,45)
\end{picture}
\put(-130,30){\vector(1,0){130}}
\put(-80,30){\circle*{1}}
\put(-80,30){\circle{2}}
\put(-30,30){\circle*{1}}
\put(-30,38){\vector(1,0){30}}
\put(-30,37){\line(0,0){2}}
\put(-80,35){\vector(1,0){50}}
\put(-30,35){\vector(-1,0){50}}
\put(-80,38){\vector(-1,0){50}}
\put(-80,37){\line(0,0){2}}
\put(-105,42){\makebox(0,0){\footnotesize{continuous series}}}
\put(-55,38){\makebox(0,0){\footnotesize{supplementary serie}}}
\put(-15,42){\makebox(0,0){\footnotesize{discrete series}}}
\put(-40,17){\makebox(0,0){\footnotesize{two mock  representations}}}
\put(-40,13){\makebox(0,0){\footnotesize{or \par $(0,0)$ representation of continuous serie}}}
\put(-75,20){\vector(-1,2){4}}
\put(-30,26){\makebox(0,0){\footnotesize{$\frac 34$}}}
\put(-82,26){\makebox(0,0){\footnotesize{$-\frac 14$}}}
\put(-115,27){\makebox(0,0){\footnotesize{$g^2$ values  }}}
\put(-75,20){\line(1,0){70}}
\put(-75,10){\line(1,0){70}}
\put(-75,10){\line(0,0){10}}
\put(-5,10){\line(0,0){10}}
\put(-25,29){\line(0,0){2}}
\put(-25,26){\makebox(0,0){\footnotesize{$\frac {15}{4}$}}}
\put(-15,29){\line(0,0){2}}
\put(-15,26){\makebox(0,0){\footnotesize{$\frac {35}{4}$}}}
\put(-7,26){\makebox(0,0){\footnotesize{$\cdots $}}}
\put(-74,5){\makebox(0,0){\footnotesize{\texttt{Figure 1}}}}
\end{center}
In all cases the full conformal symmetry is preserved.
 \par
 The paper is organized as follows. In Section 2 we remind briefly the results obtained in Refs. \cite{b33} and \cite{b37} concerning  the classical $SL(2,\mR)$ dynamics.  Section 3 deals with discrete series.  The continuous  series are considered in Section 4. Section 5, 6 and 7  are devoted to the intermediate interval $-\frac14\leq g^2\leq\frac34$ of the coupling constant values as well as  some special cases. Finally, Section 8 contains some conclusions. 
 \section{Classical $SL(2,\mR)$-invariant systems}
 The $SL(2,R)$ group is locally isomorphic  to $SO(2,1)$. This is easily seen by defining 
 \be{5}
 K=M_0+M_1,\quad  H=M_0-M_1,\quad D=M_2.
 \ee
 Then the $so(2,1)$ algebra
 \be{6}
 [M_\mu,M_\nu]=-i{\varepsilon_{\mu\nu}}^\alpha M_\alpha,
 \ee
 (we adopt  the conventions $\varepsilon_{012}=1$, $g_{\mu\nu}=\textrm{diag}(+,-,-))$ becomes 
 \be{7}
 [D,H]=-iH,\quad [D,K]=iK, \quad [K,H]=-2iD.
 \ee
Let $\xi_{\alpha}$ be the coordinates in the dual space to $so(2,1)$; the coadjoint  action of  $SL(2,\mR)$ (faithful action of $SO(2,1)$) reads 
\be{8}
\xi_\alpha'={(\Lambda^{-1})^\beta}_\alpha \xi_\beta.
\ee
The invariant (degenerate) Poisson structure reads 
\be{9}
 \{\xi_\alpha,\xi_\beta \} =-{\varepsilon_{\alpha\beta}}^\gamma\xi_\gamma.
\ee
There are three  families of coadjoint orbits:
\begin{enumerate}
\item[(i)]
upper ($\xi_0>0$) and lower ($\xi_0<0$) sheets of two-sheeted hyperboloids 
\be{10}
\xi^\alpha\xi_\alpha=\lambda^2>0,
\ee
\item[(ii)] one-sheeted hyperboloids 
\be{11}
\xi^\alpha\xi_\alpha=-\lambda^2<0,
\ee
\item[(iii)]
forward ($\xi_0>0$) and backward ($\xi_0<0$) cones
\be{12}
\xi^\alpha\xi_\alpha=0.
\ee
\end{enumerate}
 According to the general theory \cite{b47}-\cite{b51} the Poisson structure (\ref{e9}), when restricted to a coadjoint orbit, becomes nondegenerate yielding the symplectic manifold with invariant action of the $SL(2,\mR)$  group  providing thus an invariant Hamiltonian formalism. According to the eqs. (\ref{e5}) the generators  of conformal algebra are  represented by the following functions
 \be{13}
 H=\xi_0-\xi_1,\quad K=\xi_0+\xi_1,\quad D=\xi_2.
 \ee
 The Hamiltonian equations of motion take the form
 \be{14}
 \dot\xi_\alpha=\{\xi_\alpha,H\},
 \ee
and yield 
\be{15}
\dot\xi_0=-\xi_2,\quad \dot\xi_1=-\xi_2,\quad \dot \xi_2=\xi_1-\xi_0.
\ee
One arrives at regular dynamics which  remains regular when restricted to the orbit.
\par
In order to make contact with  standard form  of the Hamiltonian mechanics one looks for the Darboux coordinate. Let us first consider the case (i)   (we take the upper sheet $\xi_0>0$ for convenience). The following transformation 
\begin{align}
\label{e16}
\xi_0&=\frac{p^2}{4}+\frac{\lambda^2}{x^2}+\frac{x^2}{4},\nonumber \\
\xi_1&=-\frac{p^2}{4}-\frac{\lambda^2}{x^2}+\frac{x^2}{4},\\
\xi_2&=-\frac{1}{2}xp,\nonumber 
\end{align}
maps the upper sheet onto the half plane $0<x<\infty$, $-\infty<p<\infty$. The mapping is smooth, one-to-one  and $(x,p)$  become the Darboux  coordinates, $\{x,p\}=1$. By comparing  eqs. (\ref{e1}), (\ref{e13}) and (\ref{e16}) we find
\be{19}
g^2=4\lambda^2.
\ee
Therefore, ones  arrives at the standard form of the conformal mechanics with repelling potential.
\par
The case (ii)  is more complicated and more interesting. One has to choose at least  two maps  to cover the phase space manifold. Both provide the local  Darboux coordinates. These details are given in Ref. \cite{b33}. However, in order to make contact with the standard form of the conformal mechanics we consider  the intersection of our hyperboloid  with the plane $\xi_0+\xi_1=0$. It consists of two straight lines $\xi_0+\xi_1=0$, $\xi_2=\pm\lambda$. Consider two submanifolds
\be{20}
M_\pm=\{\xi_\alpha\xi^\alpha=-\lambda^2\ |\quad 0\lessgtr \xi_0+\xi_1\}.
\ee
Together with two lines defined above they cover the whole hyperboloid (see Figure 1 in Refs. \cite{b33,b37}). Equations 
\begin{align}
\label{e21}
\xi_0&=\frac{p^2}{4}-\frac{\lambda^2}{x^2}+\frac{x^2}{4},\nonumber\\
\xi_1&=-\frac{p^2}{4}+\frac{\lambda^2}{x^2}+\frac{x^2}{4},\\
\xi_2&=-\frac{1}{2}xp,\nonumber
\end{align}
provide the smooth one-to-one mapping of $M_+$ onto the half-plane $x>0$ , $-\infty<p<\infty$. Analogously,
\begin{align}
\label{e24}
\xi_0&=-\frac{p^2}{4}+\frac{\lambda^2}{x^2}-\frac{x^2}{4},\nonumber\\
\xi_1&=\frac{p^2}{4}-\frac{\lambda^2}{x^2}-\frac{x^2}{4},\\
\xi_2&=-\frac{1}{2}xp,\nonumber
\end{align} 
define the smooth one-to-one mapping of $M_-$ onto the half-plane  $x<0,-\infty<p<\infty$. For both mappings $(x,p)$ are Darboux coordinates. We see that the image  of $M_+$ yields the  standard form of the conformal mechanics  with
\be{27}
g^2=-4\lambda^2<0.
\ee 
However, we conclude that the description of the dynamics in terms of positive values of $x$  coordinate  is incomplete. The singularity related to the effect of the  falling on the center in finite time is spurious. It  is an  artefact of the choice of coordinates in the  symplectic manifold. The situation is somewhat similar to that encountered  in general relativity. For example, the only real singularity in Schwarzschild solution  resides at the center  while with the standard choice of the coordinates the metric  diverges at the horizon.
\par
Once the necessity of adjoing the submanifold $M_-$ is clearly  recognized the apparent singularity at the origin ($x=0$) disappears. Let us stress again  that  due to the fact that $M_+$ and $M_-$ do not cover the whole phase manifold the above singularity is still present in the explicit formulae  given above.  However, the rules relating the dynamics for $x>0$  and $x<0$ are  uniquely defined  (also on the Hamiltonian level).  Alternatively one could  work  with genuine covering of the phase manifold with no (even apparent) singularities. Finally, the case of light cones corresponds to the free motion. Let us consider, for example,  the forward cone $\xi_0>0$. Define the canonical (Darboux) variables by 
\begin{align}
\label{e28}
\xi_1&=-\frac{p^2}{4}+\frac{x^2}{4},\\
\label{e29}
\xi_2&=-\frac{1}{2}xp.
\end{align} 
Defining further
\be{30}
\xi=\xi_1+i\xi_2, \quad u=\frac 12(x-ip),
\ee
one can rewrite  eqs. (\ref{e28}) and (\ref{e29})  as
\be{31}
\xi=u^2.
\ee
Therefore, the  Darboux variables parametrize the Riemann surface of $\sqrt\xi$.
\par We conclude that, in the classical case, the coadjoint orbits method allows us to completely regular dynamics  invariant under the action of conformal $SL(2,\mR)$ group. In the following section we construct the quantum counterparts of such dynamical systems exhibiting the unbroken conformal symmetry.
\section{Quantum conformal mechanics: discrete series}
There  exist two discrete series $D_m^\pm$, $m=1,2,3,\ldots $, of the  inequivalent unitary irreducible representations of the $SL(2,\mR)$  \cite{b44}-\cite{b46}. The  representation $D_m^+$ acts in the Hilbert space of functions analytic in the upper half-plane ($z=x+iy$, $\textrm{Im} z>0$) equipped with the scalar product:
\be{32}
(f,g)=\frac{1}{\Gamma(m)}\int_{y>0}y^{m-1}\overline {f(z)}g(z)dxdy. 
\ee
The action of $g=\left (
\begin{array}{cc}
g_{11}&g_{12}\\
g_{21}&g_{22}
\end{array}
\right)\in SL(2,\mR)$ is given by
\be{33}
\left(D^+_m(g)f\right)(z)=(g_{12}z+g_{22})^{-m-1}f\left(\frac{g_{11}z+g_{21}}{g_{12}z+g_{22}}\right).
\ee
Analogously, $D_m^-$ acts in the Hilbert space of functions analytic in the lower half-plane equipped with the scalar product 
\be{34}
(f,g)=\frac{1}{\Gamma(m)}\int_{y<0}|y|^{m-1}\overline {f(z)}g(z)dxdy. 
\ee
\par 
Let us consider the $D^+_m$ serie. In order to find the relevant generators representing $\hat H, \hat K,$ and $\hat D$ we adopt the following form of the generators in the defining representation of $SL(2,\mR)$.
\be{36}
H=i\sigma_+,\quad K=-i\sigma_-,\quad D=-\frac{i}{2}\sigma_3.
\ee
Consider first the Hamiltonian $\hat H$. Due to 
\be{37}
e^{i\lambda H}=e^{-\lambda\sigma_+}=\left (
\begin{array}{cc}
1&-\lambda\\
0&1
\end{array}
\right),
\ee
we find 
\be{e38}
(e^{i\lambda\hat H}f)(z)=(1-\lambda z)^{-m-1}f\left(\frac{z}{1-\lambda z}\right).
\ee
Expanding both sides in $\lambda$ and comparing the linear terms we find
\be{39}
\hat H=-i(m+1)z-iz^2\frac{d}{dz}.
\ee 
Similarly,
\begin{equation}
\label{e40b}
\begin{split}
\hat K&=-i\frac{d}{dz},\\
\hat D&=-iz\frac{d}{dz}-i\frac{(m+1)}{2}.
\end{split}
\end{equation}
Computing the Casimir operator (\ref{e4})  yileds 
\be{41}
g^2=m^2-\frac14,\quad m=1,2,,\ldots
\ee
Therefore the coupling constant takes the values in discrete set 
\be{42}
g^2=\frac 3 4,\frac{ 15}{ 4},\ldots
\ee
(for other values of $g^2\geq \frac 34$,  see Section 7).
It is easy to find the eigenvectors of $\hat H$:
\be{43}
\hat Hf_E(z)=Ef_E(z).
\ee 
The energy spectrum is purely continuous  and positive. The eigenvectors read 
\be{44}
f_E(z)=\frac{(2E)^\frac m2}{\sqrt{2\pi}}e^{\frac {-iE}{z}}z^{-(m+1)}.
\ee 
They are orthonormal and form a complete set 
\begin{equation}
\label{e45}
\begin{split}
(f_E,f_{E'})&=\delta(E-E'),\\
\int\limits_0^\infty d E(g,f_E)(f_E,h)&=(g,h).
\end{split}
\end{equation}
In order to make contact with the picture sketched  in the previous section we need also the eigenvectors of $\hat K$: 
\be{46}
\hat K g_\lambda(z)=\lambda g_\lambda(z).
\ee 
The  spectrum is again purely continuous and positive. The eigenvectors read 
\be{47}
g_\lambda(z)=\frac{(2\lambda)^\frac m2}{\sqrt{2\pi}}e^{ {i\lambda}{z}},
\ee 
and form the orthonormal and complete set.
In order to define the coordinate representation we put
\be{48}
\lambda=\frac{x^2}{2}, \quad x>0;
\ee
and normalize  $g_\lambda(z)$  to $\delta(x-x')$; so  eq. (\ref{e47}) is replaced by
\be{49}
g_x(z)=\frac{x^{m+\frac 12}}{\sqrt {2\pi}}e^{\frac{ix^2z}{2}}.
\ee
Consequently,  the generator of the  special conformal transformations takes the form 
\be{51}
\hat K=\frac{x^2}{2}.
\ee 
Now, one can compute the wave function of any vector $f(z)$  in  the coordinate representation ($z=w+iv$)
\be{50}
\tilde f(x)=(g_x,f)=\frac{1}{\Gamma(m)}\int\limits_{-\infty}^\infty dw\int\limits_0^\infty dv v^{m-1}\overline{g_x(z)}f(z)
\ee
Let us compute the wave functions of the  Hamiltonian eigenvectors
\begin{align}
\label{e52}
\tilde f_E(x)=(g_x,f_E)=&\frac{x^{m+\frac 12}(2E)^{\frac m2}}{2\pi\Gamma(m)}\int\limits_{-\infty}^\infty dw\int\limits_0^\infty dv v^{m-1}
(w+iv)^{-(m+1)} \nonumber \\
&\cdot\exp\left(-i\left(\frac{x^2}{2}(w-iv)+\frac{E}{w+iv}\right)\right).
\end{align}
In polar coordinates (\ref{e52}) takes the form 
\be{53}
\tilde f_E(x)=\frac{x^{m+\frac 12}(2E)^{\frac m2}}{2\pi\Gamma(m)}\int\limits_{0}^\infty dr\int\limits_0^\pi d\theta r^{-1} \sin^{m-1}\theta e^{-i(m+1)\theta}
e^{-ie^{-i\theta}(\frac{x^2r}{2}+\frac{E}{r})}.
\ee
First, we do the $r$-integration (see \cite{b52})
\be{54}
\int\limits_0^\infty dr r^{-1}\exp\left(-i\left(\frac{x^2e^{-i\theta}}{2}r+\frac{Ee^{-i\theta}}{r}\right)\right)=2K_0(x\sqrt{2E}ie^{-i\theta}),
\ee
so that  eq. (\ref{e53})  takes the form 
\be{55}
\tilde f_E(x)=\frac{x^{m+\frac 12}(2E)^{\frac m2}}{\pi\Gamma(m)}\int\limits_0^\pi d\theta  \sin^{m-1}\theta e^{-i(m+1)\theta} K_0(x\sqrt{2E}ie^{-i\theta}).
\ee
Using (see again  \cite{b52})
\begin{align}
\label{e56}
K_0(z)&=-\ln\left(\frac z 2\right)I_0(z)+\sum\limits_{k=0}^\infty\frac{\psi(k+1)}{2^{2k}(k!)^2}z^{2k},\\
I_0(z)&=\sum\limits_{k=0}^\infty\frac{1}{(k!)^2}\left(\frac z 2\right)^{2k},
\end{align}
one easily concludes that the second term on the right-hand side of eq. (\ref{e56}) does not contribute. The first one yields
\be{57}
\tilde f_E(x)=i^{-(m+1)}\sqrt{x}J_m(\sqrt{2E}x),
\ee
which  proves that 
\be{58}
\hat H=-\frac 12\frac{d^2}{dx^2}+\frac{m^2-\frac 14}{2x^2}.
\ee
Operator $\hat D$ can be recovered in a similar way. 
\par
Using the isomorphism  between $SL(2,\mR)$ and $SU(1,1)$ defined by  
\be{59}
SU(1,1)\ni \left (
\begin{array}{cc}
\alpha&\beta\\
\overline\beta&\overline \alpha
\end{array}
\right)\rightarrow \left (
\begin{array}{cc}
\textrm{Re}(\alpha+\beta)&-\textrm{Im}(\alpha-\beta)\\
\textrm{Im}(\alpha+\beta)&\textrm{Re}(\alpha-\beta)
\end{array}
\right)\in SL(2,\mR),
\ee
one can construct an alternative  model of the  unitary irreducible representations of the discrete series. To this end one considers the Hilbert space of  the functions analytic in the unit disc ($w=x+iy$, $|w|<1$) and equipped with the scalar product  
\be{60}
(f,g)=\frac{1}{\Gamma(m)}\int\limits_{|w|<1}\overline{f(w)}g(w)(1-|w|^2)^{m-1}dxdy.
\ee
The unitary representation $\tilde D^+_m$ of $SU(1,1)$ (and, consequently also $SL(2,\mR)$) is given by
\be{61}
\left(\tilde D^+_m(g)f\right)(w)=(\overline\alpha+\beta w)^{-(m+1)}f\left(\frac{\alpha w+\overline \beta}{\overline\alpha+\beta w}\right).
\ee
The representation $\tilde D_m^-$ is obtained by the formula $\tilde D^-_m(g)=\tilde D^+_m(\overline g)$.
The relation between the representations expressed in terms of the analytic functions on the upper half-plane and unit disc reads
\begin{equation}
\label{e63a}
\begin{split}
&w\equiv w(z)=\frac{z-i}{z+i},\quad z\equiv z(w)=i\frac{1+w}{1-w}, \\
&\tilde f(w)=2(1-w)^{-(m+1)}f(z(w));
\end{split}
\end{equation}
here $z$ belongs to the upper half-plane while $w$  to the open unit disc. Eqs. (\ref{e63a}) and (\ref{e39})-(\ref{e40b}) lead to  the following form of generators 
\begin{equation}
\label{e64}
\begin{split}
\hat\xi_0&=\frac 12(\hat K+\hat H)=\frac{m+1}{2}+w\frac{d}{dw},\\
\hat\xi_1&=\frac 12(\hat K-\hat H)=-\frac{m+1}{2}w-\frac12(1+w^2)\frac{d}{dw},\\
\hat\xi_2&=\hat D=i\frac{m+1}{2}w-\frac{i}{2}(1-w^2)\frac{d}{dw}.
\end{split}
\end{equation}  
It is not difficult to construct, using  the above realization in terms  of analytic functions on the unit disc, the conformal dynamics in terms of wave functions  defined on phase space. To this end  let us note there is one-to-one correspondence between the points on the  unit disc and on the (upper) sheet of the unit hyperboloid
\be{65}
(\xi^0)^2-(\xi^1)^2-(\xi^2)^2=1.
\ee
It reads 
\be{66}
\xi^0=\frac{1+|w|^2}{1-|w|^2},\quad \xi^1+i\xi^2=\frac{2w}{1-|w|^2}.
\ee
Let us consider an arbitrary hyperboloid 
\be{67}
(\xi^0)^2-(\xi^1)^2-(\xi^2)^2=\lambda^2, \quad \lambda >0.
\ee
Upon rescaling $\xi^\mu\rightarrow \xi^\mu/\lambda $ one converts the manifold (\ref{e67})  into the unit hyperboloid (\ref{e65}). Therefore, the former can be parametrized  as follows
\be{68}
\xi^0=\lambda\frac{1+|w|^2}{1-|w|^2},\quad \xi^1+i\xi^2=\frac{2\lambda w}{1-|w|^2}.
\ee
It is not difficult to check that the $SU(1,1)$ action on  unit disc is equivalent  to the action of $SO(2,1)$ Lorentz group on $\xi$ variables
\be{69}
w'(\xi)=w(\xi'), \quad \xi'=\Lambda(g)\xi, \quad g \in SU(1,1).
\ee   
Let us define the function $\hat f(\xi)$  by
\be{70}
\hat f(\xi)=\frac{1}{\sqrt{\Gamma(m)}}\frac{(2\lambda)^{\frac m2}}{(\xi^0+\lambda)^{\frac{m+1}{2}}}\tilde f(w(\xi)).
\ee
Then the scalar product (\ref{e60})   becomes 
\be{71}
\int\overline{\hat  f(\xi)}\hat g(\xi)\theta(\xi^0)\delta(\xi^2-\lambda^2)d^3\xi,
\ee
while the action of $SU(1,1)\simeq  SL(2,\mR)$  takes the form
\be{72}
\left(D^+_m(g)\hat f\right)(\xi)=\left(\frac{\overline\alpha(\xi^0+\lambda)+\beta(\xi^1+i\xi^2)}{|\overline\alpha(\xi^0+\lambda)+\beta(\xi^1+i\xi^2)|}\right)^{-(m+1)}\hat f(\Lambda(g)\xi).
\ee 
However, one should take  into account that  the functions $f(w)$ are analytic. This imposes some constraints  for $\hat f(\xi)$. In terms of the functions defined on the unit disc  it is simply the Cauchy-Riemann equation $\frac{\partial f}{ \partial \overline w}=0$ which, due to (\ref{e70})  takes the form
\be{73}
\bigtriangleup \hat f(\xi)=0,
\ee
 where $\xi=\xi^1+i\xi^2$ and 
 \be{74}
 \Delta =\frac{m+1}{2}\xi+(\xi^0+\lambda)^2\frac{\partial}{\partial \overline \xi}+\xi^2\frac{\partial}{\partial \xi}.
 \ee
 The infinitesimal form of eq. (\ref{e72}) yields
  \begin{equation}
    \label{e75}
    \begin{split}
  \hat H&=i\xi^2\frac{\partial}{\partial \xi^1}-i(\xi^0+\xi^1)\frac{\xi}{\partial \xi^2}+\frac{m+1}{2}\left(1+\frac{\xi^1}{\xi^0+\lambda}\right),\\
\hat K&=i\xi^2\frac{\partial}{\partial \xi^1}+i(\xi^0-\xi^1)\frac{\xi}{\partial \xi^2}+\frac{m+1}{2}\left(1-\frac{\xi^1}{\xi^0+\lambda}\right),\\
 \hat D&=-i\xi^0\frac{\partial}{\partial\xi^1}-\frac{m+1}{2}\frac{\xi^2}{\xi^0+\lambda}.
\end{split}
\end{equation}
The invariance of the condition (\ref{e73})  follows  from the commutation rules 
\begin{equation}
\label{e76}
\begin{split}
\frac 12[\hat H+\hat K,\bigtriangleup ]&=\bigtriangleup, \\
\frac 12[\hat K-\hat H,\bigtriangleup ]&=\frac{-\overline \xi}{\xi^0+\lambda}\bigtriangleup, \\
[\hat D,\bigtriangleup ]&=\frac{ -i\xi}{\xi^0+\lambda}\bigtriangleup .
\end{split}
\end{equation}
Up to now the parameters $\lambda $ and $m$ stay unrelated; $\lambda $ sets the scale of hyperboloid on which the wave functions are supported while $m$ determines the value  of the  Casimir operator.  To relate these quantities let us remind that in the classical theory  the former parameter determines both  the size of hyperboloid  (the phase space) and  the value of Casimir (and, simultaneously, that of the coupling constant). This suggests that the desired relation between $\lambda$ and $m$ is obtained by demanding  that our  operators  (\ref{e75})     arise in the process of geometric quantization  of the classical systems constructed with the help of  the orbits method. Using   the results obtained in Ref. \cite{b53} we find that the eqs.  (\ref{e75})  (or, equivalently, eqs. (\ref{e64}))  are obtained by quantizing the classical $SL(2,\mR)$-invariant system provided 
\be{77}
\lambda =\frac{m+1}{2}.
\ee 
Let us comment on the above formula. Consider the $x$-$p$ representation of the $sl(2,\mR)$ Lie algebra. The value of the relevant Casimir operator reads (cf. eq. (\ref{e4}))
\be{79}
\hat C=\frac{g^2}{4}-\frac {3} {16}.
\ee 
The representation (\ref{e75}) of $sl(2,\mR)$ yields
\be{80}
\hat C=\frac{m^2-1}{4}.
\ee
Comparing the above expressions for $\hat C$ we get the relation between $g$  nad $m$  
\be{81}
g^2=m^2-\frac 14.
\ee
On the other hand  $g$ is related to the size of hyperboloid by eq. (\ref{e19}) :
\be{82}
g^2=4\lambda^2=4((\xi_0)^2-(\xi_1)^2-(\xi_2)^2).
\ee
Quantizing $\xi_\mu$, $\xi_\mu\rightarrow \hat\xi_\mu$ we find (see \cite{b53})
\be{83}
(\hat \xi_0)^2-(\hat \xi_1)^2-(\hat \xi_2)^2=\lambda (\lambda -1),
\ee
which, by (\ref{e77}), agrees with (\ref{e80}). The minimal value of $\lambda$ is $1$; then $C=0$ and $g^2=\frac 34$. The "classical" phase space is then the (upper) sheet of the  hyperboloid 
\be{84}
(\xi_0)^2-(\xi_1)^2-(\xi_2)^2=1.
\ee 
However, we should keep in mind  that we are using the units with $\hbar =1$. Reinserting $\hbar$  one concludes that the size of  the classical phase space  would be of order $\hbar$. The genuine  classical limit is obtained by taking $\hbar\rightarrow 0$, $m\rightarrow \infty$ and $m\hbar=\textrm{const}$.
\par 
In a similar way  one can analyse the case of lower sheet  and/or the second discrete  serie $D^-_m$.
\section{Quantum conformal mechanics: continuous series}
The $SL(2,\mR)\simeq SU(1,1)$    group possesses the continuous series of  the  unitary  irreducible  representations \cite{b44}-\cite{b46}. They can be described as follows: the relevant Hilbert space consists of functions defined on the unit circle and square  integrable  with respect to the standard Lebesgue measure. The action of the $SU(1,1)$ group is determined  by  two parameters, $\rho\in \mR$ and $\epsilon=0,1$ and is given by
\be{78}
\left(\tilde D^{\rho,\epsilon}(g)f)\right)(e^{i\psi})=|\overline\alpha+\beta e^{i\psi}|^{i\rho-1-\epsilon }(\alpha+\overline \beta e^{-i\psi})^\epsilon f\left(\frac{\alpha e^{i\psi}+\overline\beta}{\overline\alpha+\beta e^{i\psi}}\right).
\ee

\par
All representations $\tilde D^{\rho,\epsilon}$ are irreducible  expect $(\rho,\epsilon)=(0,1)$ when we are dealing with the sum of  two irreducible representations. Two representations $(\rho,\epsilon)$ and $(\rho',\epsilon')$ are equivalent  if and only if $\rho=\rho'$, $\epsilon=\epsilon'$ or $\rho=-\rho'$, $\epsilon=\epsilon'$.  In this section we will be  dealing with the representations corresponding to $\rho\neq0$.  Therefore, we can assume $\rho>0$, $\epsilon=0,1$. The corresponding  generators  are easily  obtained  via  the isomorphism (\ref{e59}). Another equivalent and convenient  form of the representation is given by the formula
\be{85}
\left( D^{\rho,\epsilon}(g)f\right)(z)=|g_{12}z+g_{22}|^{i\rho-1-\epsilon }(g_{12}z+g_{22})^\epsilon f\left(\frac{g_{11}z+g_{21}}{g_{12}z+g_{22}}\right),
\ee
where $f\in L^2(\mR)$, again with respect to the standard Lebesgue measure.
\par
As in the case of discrete series we used  (\ref{e36}) to find the representation  of the Lie algebra generators. However, to make contact with the classical theory  presented  in Section 2 it is more convenient  to use the equivalent  representation:
\be{86}
H=-i\sigma_+,\quad K=i\sigma_-,\quad D=-\frac{i}{2}\sigma_3.
\ee
This choice leads to the following form of generators 
\begin{equation}
\label{e87}
\begin{split}
\hat H&=-\frac{\rho+i}{2}\sin\psi-\frac\epsilon 2 (1+\cos\psi)+i(1+\cos\psi)\frac{d}{d\psi},\\
\hat K&=\frac{\rho+i}{2}\sin\psi-\frac\epsilon 2 (1-\cos\psi)+i(1-\cos\psi)\frac{d}{d\psi},\\
\hat D&=\frac{\rho+i}{2}\cos\psi-\frac\epsilon 2 \sin\psi+i\sin\psi\frac{d}{d\psi}.
\end{split}
\end{equation}
Computing the Casimir operator yields
\be{88}
\hat C=-\frac 14(\rho^2+1)<-\frac14.
\ee
Our next step is to find the spectrum of the Hamiltonian $\hat H$ and the generator $\hat K$ of the special conformal transformation. 
The solutions to the eigenvalue equation 
\be{89}
\hat H f_E(\psi)=Ef_E(\psi),
\ee
reads
\be{90}
f_E(\psi)=\frac{1}{\sqrt{2\pi}}e^{\frac{-i\epsilon\psi}{2}}e^{-iE\tan(\frac\psi 2)}(1+\cos\psi)^{\frac{i\rho-1}{2}}\left\{
\begin{array}{c}
1,\quad 0\leq \psi\leq\pi,\\
(-1)^\epsilon,\quad \pi< \psi<2\pi.
\end{array} 
\right.
\ee
The eigenfunctions $f_E(\psi)$ are properly normalized 
\be{91}
(f_E,f_{E'})=\delta(E-E'),
\ee
 and obey the completeness relation 
\be{92}
\int\limits_{-\infty}^\infty dEf_E(\psi)\overline{f_E(\psi')}=\delta(\psi-\psi').
\ee 
Similarly, one can look for the eigenvalue problem for the conformal generator $\hat K$
\be{93}
\hat K g_\kappa(\psi)=\kappa g_\kappa(\psi).
\ee 
The solutions to eq. (\ref{e93}) read
\be{94}
g_\kappa(\psi)=\frac{1}{\sqrt{2\pi}}e^{\frac{-i\epsilon\psi}{2}}e^{i\kappa\cot(\frac\psi 2)}(1-\cos\psi)^{\frac{i\rho-1}{2}}.
\ee
Again, they are normalized to  $\delta(\kappa-\kappa')$ and obey the completeness  relation if the integration over $\kappa$ extends over the whole real axis.
\par
We conclude that both $\hat H$ and $\hat K$ have purely continuous spectrum extending from $-\infty$ to $\infty$. Comparing the energy spectrum  with the properties of the classical motion  we conclude that the continuous corresponds to the motion on one-sheeted hyperboloid.
\par
In order to make contact with the classical parametrization (\ref{e21}) and (\ref{e24}) we parametrize  its spectrum as follows 
\be{95}
\kappa=\left\{
\begin{array}{c}
\frac{x^2}{2},\quad x>0,\\
-\frac{x^2}{2}, \quad x<0.
\end{array}
\right.
\ee  
The new representation  with $\hat K$ diagonal  is defined as 
\be{96}
\tilde f(x)=|x|^{\frac 12+i\rho}(g_\kappa,f),
\ee
with $\kappa$ being  related to $x$ via formula (\ref{e95}). The prefactor $|x|^\frac 12$ is introduced in order to  obtain  the Hilbert space of functions square integrable with respect to the standard  Lebesque measure on $\mR$. An additional phase  factor $|x| ^{i\rho}$ is added to provide  the proper form of generators in $x$-representation.
\par
Explicitly, we have 
\be{97}
\tilde f(x)=\frac{|x|^{\frac 12+ i\rho}}{\sqrt{2\pi}}\left\{
\begin{array}{c}
\int\limits_0^{2\pi}d\psi e^{\frac{i\epsilon\psi}{2}}e^{\frac{-ix^2}{2}\cot(\frac\psi 2)}(1-\cos\psi)^{-\frac{i\rho+1}{2}}f(\psi),\quad  x>0,\\
\int\limits_0^{2\pi}d\psi e^{\frac{i\epsilon\psi}{2}}e^{\frac{ix^2}{2}\cot(\frac\psi 2)}(1-\cos\psi)^{-\frac{i\rho+1}{2}}f(\psi),\quad  x<0.
\end{array}
\right.
\ee
In particular, it is not difficult to find the energy eigenfunctions in the $x$ representation. Using eq. (\ref{e90}) we find 
\begin{align}
\label{98}
\tilde f_E(x)&=|x|^{\frac 12+i\rho}(g_\kappa,f_E)=|x|^{\frac 12+i\rho}\int\limits_0^{2\pi}\overline{g_\kappa(\psi)}f_E(\psi)\nonumber \\
&=\frac{|x|^{\frac 12+i\rho}}{2\pi}\int\limits_0^\pi d\psi e^{-i\kappa\cot(\frac\psi2) -iE\tan(\frac\psi2)}(1+\cos\psi)^\frac{i\rho-1}{2}(1-\cos\psi)^\frac{-i\rho-1}{2}\nonumber \\
&+\frac{(-1)^\epsilon|x|^{\frac 12+i\rho}}{2\pi}\int\limits_\pi^{2\pi} d\psi e^{-i\kappa\cot(\frac\psi2) -iE\tan(\frac\psi2)}(1+\cos\psi)^\frac{i\rho-1}{2}(1-\cos\psi)^\frac{-i\rho-1}{2},
\end{align}
where $\kappa$ is related to $x$ by eq. (\ref{e95}). Making the change of variables $u=\tan(\frac\psi 2)$, one finds after some manipulations
\be{99}
\tilde f_E(x)=\left\{
\begin{array}{c}
\frac{|x|^{\frac 12+i\rho}}{\pi}\int\limits_0^\infty du u^{-i\rho-1}\cos(Eu+\frac\kappa u), \quad \epsilon=0,\\
\frac{-i|x|^{\frac 12+i\rho}}{\pi}\int\limits_0^\infty du u^{-i\rho-1}\sin(Eu+\frac\kappa u), \quad \epsilon=1.
\end{array}
\right.
\ee
The above integrals can be easily taken (see \cite{b52}) to yield:
\\
\par -- for $\epsilon=0$, $Ex>0$
\begin{align}
\label{e100}
\tilde f_E(x)=(2|E|)^{\frac{i\rho}{2}}|x|^{\frac 12}\left(\right. & iJ_{-i\rho}(|x|\sqrt{2|E|})\sinh(\frac{\pi\rho}{2})\nonumber\\
&\left. -N_{-i\rho}(|x|\sqrt{2|E|})\cosh(\frac{\pi\rho}{2})\right);
\end{align}
\par --  for $\epsilon=0$, $Ex<0$
\be{100b}
\tilde f_E(x)=\frac{-2}{\pi}(2|E|)^{\frac{i\rho}{2}}|x|^{\frac 12}K_{-i\rho}(|x|\sqrt{2|E|})\cosh(\frac{\pi\rho}{2});
\ee
\par --  for $\epsilon=1$, $Ex>0$
\begin{align}
\label{100c}
\tilde f_E(x)=\textrm{sgn} (E)(2|E|)^{\frac{i\rho}{2}}|x|^{\frac 12}\left(\right.&-iJ_{-i\rho}(|x|\sqrt{2|E|})\cosh(\frac{\pi\rho}{2})\nonumber\\
&\left. +N_{-i\rho}(|x|\sqrt{2|E|})\sinh(\frac{\pi\rho}{2})\right);
\end{align}
\par -- for $\epsilon=1$, $Ex<0$
\be{100d}
\tilde f_E(x)=\frac{-2\textrm{sgn} (E)}{\pi}(2|E|)^{\frac{i\rho}{2}}|x|^{\frac 12}K_{-i\rho}(|x|\sqrt{2|E|})\sinh(\frac{\pi\rho}{2}).
\ee
We see that  $\tilde f_E(x)$ obey the eigenvalue equation(s):
\begin{equation}
\label{e101}
\begin{split}
\left(-\frac 12\frac{d^2}{dx^2}-\frac{\rho^2+\frac 14}{2x^2}\right)\tilde f_E(x)=E\tilde f_E(x), \quad x>0;\\
\left(\frac 12\frac{d^2}{dx^2}+\frac{\rho^2+\frac 14}{2x^2}\right)\tilde f_E(x)=E\tilde f_E(x), \quad x<0;
\end{split}
\end{equation}
The relation between the parameter $\rho$ and the coupling constant $g$ (cf. eq. (\ref{e1})) reads 
\be{102}
g^2=-(\rho^2+\frac 14)=(i\rho)^2-\frac 14.
\ee
Eqs. (\ref{e100})-(\ref{e100d}) determine the way  the solutions of the eigenvalue equations for $x>0$ and $x<0$ are glued together at $x=0$ to yield the eigenvectors of the  self-adjoint generator $\hat H$. If we restrict  ourselves  to the $x>0$ region for negative  coupling constant then the  conformal symmetry remains unbroken. This is not possible for eigenvalue  problem on semiaxis where any boundary condition defining a self-adjoint extension  of the symmetric operator  given by the formal differential  expression $-\frac 12\frac{d^2}{x^2}-\frac{g^2}{2x^2}$ breaks the conformal symmetry \cite{b42}.  
\par
It remains to relate the formalism presented above to the one emerging  from canonical (geometric) quantization of the Hamiltonian dynamics defined on the one-sheeted hyperboloid. This hyperboloid is a submanifold of the three-dimensional  space carrying  the linear representation of $SO(2,1)\simeq SL(2,\mR)/\mathbb{Z}_2$ (the adjoint representation $SO(2,1) $ or $SL(2,\mR)$). On the other hand, as it is clearly seen from eq. (\ref{e78}), the irreducible  representations  under considerations are spanned by the function supported  on compact manifold ($S^1$) on which $SL(2,\mR)$ acts nonlinearity. This action is transitive so it  results from  nonlinear action of the group on compact coset manifold. Consider the compact subgroup generated by $H+K$. The relevant  group manifold can be viewed as the coset manifold $SL(2,\mR)/G(D,K)$ where $G(D,K)\subset SL(2,\mR)$  is the subgroup  generated by $D$ and $K$. With our choice  (\ref{e86}) 
\be{103}
H+K=\sigma_2.
\ee
The nonlinear action of $g=\left (
\begin{array}{cc}
a&b\\
c&d
\end{array}
\right)\in SL(2,\mR)$ , $ad-bc=1$, is given by 
\be{104}
ge^{i\theta(H+K)}=e^{i\theta'(H+K)}e^{i\gamma(g,\theta)D}e^{i\delta(
g,\theta)K}.
\ee
Eq. (\ref{e104}) implies 
\begin{equation}
\label{e105}
\begin{split}
\tan \theta'&=\frac{a\tan\theta+b}{c\tan\theta+d},\\
e^\gamma &=\frac{1+\tan^2\theta}{(a\tan\theta+b)^2+(c\tan \theta+d)^2},\\
\delta&=\frac{(b^2+d^2-a^2-c^2)\tan\theta+(ab+cd)(\tan^2\theta-1)}{(a\tan\theta+b)^2+(c\tan \theta+d)^2}.
\end{split}
\end{equation}
The action of $SU(1,1)$ on the variable $\psi$ following from eq. (\ref{e78})  reads 
\be{106}
e^{i\psi'}=\frac{\overline\alpha e^{i\psi}-\overline\beta}{\alpha-\beta e^{i\psi}}.
\ee
 By comparing  eqs. (\ref{e105}) and (\ref{e106}) and using the isomorphism (\ref{e59}) we conclude that one can identify
 \be{107}
 \psi=2\theta.
 \ee
 In order to construct the variables transformation  according to the linear  representation of $SL(2,\mR)$  we follow  the method described  in Refs. \cite{b54,b54b}. First, we construct  the three-dimensional  representation $\mathcal{D}$  of $SL(2,\mR)$  -- the adjoint representation:
 \begin{align}
 \label{e108}
 \mathcal{D}(e^{i\lambda D})=\left (
\begin{array}{ccc}
\cosh\lambda&-\sinh\lambda&0\\
-\sinh\lambda&\cos\lambda&0\\
0&0&1
\end{array}
\right),
\mathcal{D}(e^{i\lambda H})=\left (
\begin{array}{ccc}
1+\frac{\lambda^2}{2}&-\frac{\lambda^2}{ 2}&-\lambda\\
\frac{\lambda^2}{2}&1-\frac{\lambda^2}{2}&-\lambda\\
-\lambda&\lambda&1
\end{array}
\right),
\\
\label{e108a}
\mathcal{D}(e^{i\lambda K})=\left (
\begin{array}{ccc}
1+\frac{\lambda^2}{2}&\frac{\lambda^2}{ 2}&\lambda\\
-\frac{\lambda^2}{2}&1-\frac{\lambda^2}{2}&-\lambda\\
\lambda&\lambda&1
\end{array}
\right), 
\mathcal{D}(e^{i\theta(H+K)})=\left (
\begin{array}{ccc}
1&0&0\\
0&\cos2\theta&-\sin2\theta\\
0 &\sin2\theta&\cos2\theta
\end{array}
\right).
 \end{align}
 The  above representation, when restricted  to the subgroup generated by $D$ and $K$ is not completely reducible. It has  the two-dimensional invariant subspace  spanned by the vectors 
 \be{109}
e_1=\left (
\begin{array}{c}
1\\-1\\0
\end{array}
\right),\quad 
e_2=\left (
\begin{array}{c}
1\\-1\\1
\end{array}
\right).
\ee
Any element of this subspace can be written  as $\chi_1e_1+\chi_2e_2$. The action of the subgroup generated by $K$ and $D$ reads\
\begin{equation}
\label{e110}
\begin{split}
e^{i\lambda K}&:\quad \chi_1'=\chi_1+\lambda\chi_2, \quad \chi_2'=\chi_2;\\
e^{i\lambda D}&:\quad \chi_1'=e^{-\lambda}\chi_1+(e^{-\lambda}-1)\chi_2, \quad \chi_2'=\chi_2;
\end{split}
\end{equation}
Now, according to the formalism described in Refs. \cite{b54,b54b}, the relation between  the parameters $\theta,\chi_1$ and $\chi_2$  transforming nonlinearly under $SL(2,\mR)$ and the coordinates $\xi_0,\xi_1,\xi_2$, transforming  linearly reads 
\be{e111}
\left (
\begin{array}{c}
\xi_0\\ \xi_1\\ \xi_2
\end{array}
\right)
=\mathcal{D}(e^{i\theta(H+K)})\left (
\begin{array}{c}
\chi_1+\chi_2\\-\chi_1-\chi_2\\ \chi_2
\end{array}
\right).
\ee
Due to (\ref{e108a})  we have the following relations
\begin{align}
\label{e112}
\xi_0&=\chi_1+\chi_2, \nonumber\\
\xi_1&=-(\chi_1+\chi_2)\cos2\theta-\chi_2\sin2\theta,\\
\xi_2&=-(\chi_1+\chi_2)\sin2\theta+\chi_2\cos2\theta \nonumber.
\end{align}
Note that 
\be{113}
(\xi_0)^2-(\xi_1)^2-(\xi_2)^2=-\chi_2^2,
\ee
which is invariant under $SO(2,1)$ as it follows from eqs. (\ref{e110}).  Taking into account  eq. (\ref{e107}) one can rewrite eq. (\ref{e112})  as 
\begin{equation}
\label{e114}
\begin{split}
\xi_1&=-\xi_0\cos\psi-\chi_2\sin\psi,\\
\xi_2&=-\xi_0\sin\psi+\chi_2\cos\psi.
\end{split}
\end{equation}
By comparing eqs. (\ref{e11}) and (\ref{e113}) we conclude that the classical phase space may parametrized by $\xi_0$ and $\psi$ with $\chi_2=\lambda$ kept fixed. The Poisson brackets (\ref{e9})  are equivalent to
\be{e115}
\{\psi,\xi_0\}=1;
\ee 
The (geometric) quantization procedure  is therefore, straightforward. On the quantum level 
\be{116}
[\hat \psi,\hat \xi_0]=i.
\ee
Let us consider the following representation  of the above commutation rule 
\be{117}
\hat \psi=\psi,\quad \hat \xi_0=-i\frac{d}{d\psi}+\frac \epsilon 2.
\ee
With the appropriate ordering rule, 
$\xi_0f(\psi)\rightarrow \frac12(\hat\xi_0f(\hat\psi)+f(\hat\psi)\hat\xi_0)$ we find from eqs. (\ref{e13}) and (\ref{e114})
\begin{equation}
\label{e118}
\begin{split}
\hat H&=(\chi_2+\frac i2)\sin\psi+\frac\epsilon 2(1+\cos\psi)-i(1+\cos\psi)\frac{d}{d\psi},\\
\hat K&=-(\chi_2+\frac i2)\sin\psi+\frac\epsilon 2(1-\cos\psi)-i(1-\cos\psi)\frac{d}{d\psi},\\
\hat D&=(\chi_2+\frac i2)\cos\psi-\frac\epsilon 2\sin\psi+\sin\psi\frac{d}{d\psi}.
\end{split}
\end{equation}
Eqs. (\ref{e118}) coincide up to the Lie algebra isomorphism  $H\rightarrow -H, K\rightarrow - K,D\rightarrow D$ with eqs. (\ref{e87}) provided the identification $\rho=2\chi_2$ has  been made. Noting that $\chi_2=\lambda$ we find from eq. (\ref{e102}) 
\be{119}
 g^2=-4\lambda^2-\frac 1 4,
\ee
which should be compared with the classical relation (\ref{e27}). We see  that these equations coincide up to the  quantum correction.
Let us note that, the parametrization (\ref{e112}) has been used by Plyushchay in his paper on quantization of $SL(2,\mR)$ symmetry \cite{b53}.
 \par
The value of the Casimir operator  is given by  eq. (\ref{e88}).
\be{119a}
\hat C=-\frac{\rho^2}{4}-\frac{1}{4}.
\ee
 Therefore, due to the eq. (\ref{e4}) we find  
\be{120}
g^2=-\rho^2-\frac 14,
\ee
so that  $g^2<-\frac 14$.
On the other hand the parametrization (\ref{e112})  yield, upon quantization, the proper value of the Casimir operator provided $\chi_2=\lambda=\rho /2$. Due to eq. (\ref{e119a})  the quantization of the one-sheeted hyperboloid $(\xi_0)^2-(\xi_1)^2-(\xi_2)^2=-\lambda ^2$ yields  
\be{121}
(\hat \xi_0)^2-(\hat \xi_1)^2-(\hat \xi_2)^2=-\lambda ^2-\frac 14.
\ee
The continuous  series correspond to the quantization of the  "classical" theories described by the phase spaces in form of one-sheeted  hyperboloids. Again, genuine  classical limit is attained by $\hbar\rightarrow 0,\rho\rightarrow \infty$ and $\rho\hbar=\textrm{constant}$. The upper limit for the coupling constant is $-\frac 14$. 
\section{Quantum conformal mechanics: supplementary serie}
The representations considered up to now, i.e., the discrete  and continuous series, cover  the whole range of coupling constant  except the interval $\langle-\frac 14,\frac 34)$. In the present section we consider the supplementary  serie  which corresponds to the interval $(-\frac 14,\frac 34)$. The special case $g^2=-\frac{1}{4}$ will be dealt with in  the next section.  
\par The supplementary serie  of the irreducible unitary representations  is defined as follows  \cite{b44}-\cite{b46}. For  any $g$ such that $0<|\rho|<1$ we define the  Hilbert space of functions  of real variable  running  over the whole real  axis $\mR$  equipped with the scalar product
\be{122}
(f,g)=\frac{1}{\Gamma(\rho)}\int\int_{\mR^2}|y_1-y_2|^{\rho-1}\overline{f(y_1)}g(y_2) d x_1 d x_2.
\ee 
The action of $SL(2,\mR)$ group  is given by the formula
\be{123}
(\Delta_\rho(g)f)(x)=|g_{12}y+g_{22}|^{-\rho-1}f\left(\frac{g_{11}y+g_{21}}{g_{12}y+g_{22}}\right).
\ee
Alternatively, one can consider the equivalent  representation of $SU(1,1)$ acting in the space of functions defined on the unit   circle equipped with the scalar product 
\be{124}
(f,g)=\frac{1}{\Gamma(\rho)}\int\int_{S^1\times S^1}\left|\sin\left(\frac{\psi_1-\psi_2}{2}\right)\right|^{\rho-1}\overline{f(\psi_1)}g(\psi_2)d \psi_1 d\psi_2.
\ee 
Then the group action reads 
\be{125}
(\tilde\Delta_\rho(g)f)(e^{i\psi})=|\overline\alpha+\beta e^{i\psi}|^{-\rho-1}f\left(\frac{\alpha e^{i\psi}+\overline \beta}{\overline \alpha+\beta e^{i\psi}}\right).
\ee
The representation $\Delta_\rho$ and $\Delta_{-\rho}$ are unitary equivalent \cite{b45,b46}. Therefore, we may restrict ourselves to  the parameters $\rho$  obeying $0<\rho <1$. In  what follows  we use  the form of representations  described by eqs.  (\ref{e122}) and (\ref{e123}). As in the  previous cases we start  with determining the  form of generators. They read 
\begin{equation}
\label{e126}
\begin{split}
\hat H&=i(\rho+1)y+iy^2\frac{d}{dy},\\
\hat K&=i\frac{d}{dy},\\
\hat D&=-i\frac{\rho+1}{2}-iy\frac{d}{dy},
\end{split}
\end{equation}  
It is again easy  to solve the spectral problems  for $\hat H$ and $\hat K$. The general solution to the eigenvalue equation
\be{127}
\hat H f_E(y)=Ef_E(y),
\ee
reads
\be{128}
f_E(y)=\frac{|E|^{\frac \rho 2}}{\sqrt{2\cos(\frac{\pi\rho}{2}})}|y|^{-(\rho+1)}e^{\frac {iE}{y}}.
\ee
The eigenfunctions $f_E(y)$ are normalized to $\delta(E-E')$  with respect to the scalar product (\ref{e122}). The completeness condition reads 
\be{129}
\int\limits_{-\infty}^\infty dE\int\int _{\mR^2}dy_1dy_2|z_1-y_1|^{\rho-1}f_E(y_1)\overline{f_E(y_2)}|y_2-z_2|^{\rho-1}=|z_1-z_2|^{\rho-1}.
\ee  
With  some effort  one can verify the above equality for $f_E(y)$ given by  eq. (\ref{e128}) (to this end  one can use the fact that the Fourier transform of the product equals the convolution of Fourier transforms of the factors as well as the form of Fourier transform  of the distribution $|y|^\lambda$, see \cite{b55}).
\par 
The eigenvalue problem for $\hat K$, 
\be{130}
\hat K g_\kappa(y)=\kappa g_\kappa(y),
\ee
can be also  easily solved 
\be{131}
g_\kappa(y)=\frac{|\kappa|^{\frac \rho 2}}{\sqrt{2\cos(\frac{\pi\rho}{2}})}e^{ {-i\kappa}{y}}.
\ee  
Again  the eigenfunctions $g_\kappa(y)$ are normalized to $\delta(\kappa-\kappa')$ (with respect to the scalar product (\ref{e122}))  and obey the completeness condition
 \be{132}
\int\limits_{-\infty}^\infty d\kappa \int\int _{\mR^2}dy_1dy_2|z_1-y_1|^{\rho-1}g_\kappa(y_1)\overline{g_\kappa(y_2)}|y_2-z_2|^{\rho-1}=|z_1-z_2|^{\rho-1}.
\ee   
We conclude that both $\hat H$ and $\hat K$ has purely continuous spectrum  extending from $-\infty$ to $\infty$. In order to construct  the "coordinate" representation we again label the eigenvalues of $\hat K$  according to eq. (\ref{e95}). The wave functions in the coordinate representation are  defined  as
\be{133}
\tilde f(x)=|x|^\frac 12(g_\kappa,f), \quad \kappa=\textrm{sgn}(x)\frac {x^2}{2}.
\ee
In  the coordinate representation $\hat K$ takes the "standard" form 
\be{134}
\hat K=\left\{
\begin{array}{c}
\frac{x^2}{2}, \quad x>0;\\
-\frac{x^2}{2}, \quad x<0.\\
\end{array}
\right.
\ee
The next step is to find the energy eigenvectors  in coordinate representation. According to the definition (\ref{e133}) we have
\be{135}
\tilde f_E(x)=\frac{|x|^{\frac 12}}{\Gamma(\rho)}\int\int_{\mR^2}dy_1dy_2|y_1-y_2|^{\rho-1}\frac{|\frac{x^2}{2}|^{\frac \rho 2}e^{i\kappa y_1}}{{2\cos(\frac{\pi\rho}{2}})}|y_2|^{-(\rho+1)}|E|^{\frac\rho 2}e^{\frac {iE}{y_2}}.
\ee
Changing the  variables $y_1-y_2\rightarrow y_1$, $y_2\rightarrow y_2$ we find
 \be{136}
\tilde f_E(x)=\frac{2^{-\frac{\rho}{2}}|x|^{\frac 12+\rho}|E|^{\frac \rho 2}}{2\Gamma(\rho)\cos(\frac{\pi\rho}{2})}\int\limits_{-\infty}^\infty dy_1|y_1|^{\rho-1}e^{i\kappa y_1}\int\limits_{-\infty}^\infty dy_2|y_2|^{-(\rho+1)}e^{i(\kappa y_2+\frac {E}{ y_2})}.
\ee
The first integral on the right hand side is the Fourier transform of the generalized function $|y|^{\rho-1}$  (see, \cite{b55})  while  the second can be  found in \cite{b52}. The final result reads 
\be{137}
\tilde f_E(x)=2\pi|x|^{\frac 12}\left(J_{-\rho}(|x|\sqrt{2|E|})\sin(\frac{\pi\rho}{2})-N_{-\rho}(|x|\sqrt{2|E|})\cos(\frac{\pi\rho}{2})\right),
\ee
for $E/x>0$ and 
\be{138}
\tilde f_E(x)=4|x|^{\frac 12}K_{-\rho}(|x|\sqrt{2|E|})\cos(\frac{\pi\rho}{2}),
\ee
for $E/x<0$. 
\par It is now easy to find the Hamiltonian in the coordinate representation. Eqs. (\ref{e137}) and (\ref{e138})  imply the following  differential equations for $\tilde f_E(x)$:
\be{139}
\left(-\frac 12\frac{d^2}{dx^2}+\frac{\rho^2-\frac 14}{2x^2}\right)\tilde f_E(x)=
 \left\{ 
\begin{array}{c}
E\tilde f_E(x), \quad x>0,\\
-E\tilde f_E(x), \quad  x<0.
\end{array}
\right.
\ee
Therefore, the Hamiltonian in coordinate  representation reads 
 \be{140}
\hat H=
 \left\{ 
\begin{array}{c}
-\frac 12\frac{d^2}{dx^2}+\frac{\rho^2-\frac 14}{2x^2}, \quad x>0,\\
\frac 12\frac{d^2}{dx^2}-\frac{\rho^2-\frac 14}{2x^2}, \quad  x<0.
\end{array}
\right.
\ee
Let us note that the coupling constant variables are in the interval $\frac{-1}{4}<g^2<\frac 34$, i.e.,  it can attain both  positive and negative values. In the classical case this  would correspond  both  to the case  of one- and two-sheeted  hyperboloids (and a light cone) as  the phase spaces. In spite of that  in the quantum case  the range of coordinate variable extends over the whole  real axis  for all $0<\rho<1$. The reason for  that is  quite  simple  and  would be clearly  seen if we introduced explicitly   the Planck constant $\hbar$.  The standard way  of taking  the classical limit is to keep the coupling fixed  (or, at least, not infinitesimally small) when  going $\hbar\rightarrow 0$.  However, this implies  the Casimir  eigenvalue tends to infinity  as $\hbar^{-2}$. This is here impossible due to the restriction $0<\rho <1$.  In other words, our potential  is proportional to $\hbar^2 $  so we  always  dealing  with purely quantum case. For example,  for classical repelling potential  we can argue that it is sufficient to restrict ourselves to the positive semiaxis. This is because the semiclassical tunneling factor $\exp(\frac{-1}{\hbar}\int^x\sqrt{V(x)-E} dx)$ goes  to zero as $x\rightarrow 0^+$. This is, however, not the case  if $V(x)$ itself is proportional  to $\hbar ^2$; in fact, the semiclassical approximation makes here no sense in  its standard form.
\par
The most interesting case corresponds to $\rho=\frac 12$. As we see from eq. (\ref{e140}) the dynamics is then given, up to sign, by  the free dynamics  on each  semiaxis. The generators of $SL(2,\mR)$  take the form
\be{141}
\hat H=
 \left\{ 
\begin{array}{c}
-\frac 12\frac{d^2}{dx^2}, \quad x>0,\\
\frac 12\frac{d^2}{dx^2}, \quad  x<0;
\end{array}
\right.
\ee
\be{142}
\hat K=
 \left\{ 
\begin{array}{c}
\frac{x^2}{2}, \quad x>0,\\
-\frac{x^2}{2}, \quad  x<0;
\end{array}
\right.
\ee
\be{143}
\hat D=\frac{i}{2}x\frac{d}{dx}+\frac i4.
\ee
It is not difficult to find the energy eigenfunctions in the coordinate  representations. Following the same way as in the previous sections we find
\be{144}
\tilde f_E(x)=\left\{
\begin{array}{c}
-2\sqrt{\frac{2\pi}{\sqrt{2|E|}}}\sin(\sqrt{2|E|}x -\frac \pi 4), \quad \frac{E}{x}>0;\\
\sqrt{2}\sqrt{\frac{2\pi}{\sqrt{2|E|}}}e^{-\sqrt{2|E|}x }, \quad \frac{E}{x}<0.
\end{array}
\right.
\ee
It is not difficult to find the global action of $\hat K$ and $\hat D$. The global action of $\hat H$ is nonlocal. Its kernel  can be be explicitly found in terms of Fresnel integrals. It is, however, not very  enlightening  so we skip  it here. It is important to note that  the case  under  consideration does not  correspond to the free motion. For example,  the spectrum of $\hat H$ as well as $\hat K$ are not bounded  from below as in the free case.
\section{Quantum conformal mechanics: the exceptional cases}
As far the unitary irreducible representations of $SL(2,\mR)$ are  concerned we are left with the case $\rho=0$, $\epsilon=0,1$. The case $\rho=0,\epsilon =0$  corresponds  to a single irreducible  representation. The case $\epsilon=1$ is more  interesting. The relevant  representation  is reducible, being direct  sum of irreducible  components. They  can be formally viewed as the limits $m\rightarrow 0^+$ of the discrete  series $D_m^\pm$.      
\par
Let us start with the case $\rho=0,\epsilon=0$.  Eq. (\ref{e78})  takes the form
\be{145}
\left(\tilde D^{0,0}(g)f\right)(e^{i\psi})=|\overline\alpha+\beta e^{i\psi}|^{-1}f\left(\frac{\alpha e^{i\psi}+\overline\beta}{\overline\alpha+\beta e^{i\psi}}\right).
\ee
 The generators can be read off from  eqs. (\ref{e87}):
 \begin{align}
\label{e146}
\hat H&=-\frac{i}{2}\sin\psi+i(1+\cos\psi)\frac{d}{d\psi},\\
\label{e146a}
\hat K&=\frac{i}{2}\sin\psi+i(1-\cos\psi)\frac{d}{d\psi},\\
\label{e146b}
\hat D&=\frac{i}{2}\cos\psi+i\sin\psi\frac{d}{d\psi},
\end{align}
which leads to the energy  eigenfunctions
\be{147}
f_E(\psi)=\frac{1}{\sqrt{2\pi}}e^{-iE\tan(\frac\psi 2)}(1+\cos \psi)^{-\frac 12},
\ee
and the eigenfunctions of $\hat K$
\be{148}
g_\kappa(\psi)=\frac{1}{\sqrt{2\pi}}e^{i\kappa\cot(\frac\psi 2)}(1-\cos \psi)^{-\frac 12}.
\ee
We adopt eq. (\ref{e96}) as defining the coordinate representation
\be{149}
\tilde f(x)=|x|^{\frac 12}(g_\kappa,f),
\ee
which leads to 
\be{149b}
\tilde f _E(x)=\left\{
\begin{array}{c}
-|x|^\frac 12 N_0(\sqrt{2|E|}|x|), \quad E/x>0;\\
-\frac {2}{\pi} |x|^\frac 12 K_0(\sqrt{2|E|}|x|), \quad E/x<0.
\end{array}
\right.
\ee
Consequently, the Hamiltonian can be  obtained by putting $\rho=0$ in eqs. (\ref{e101}).  
\par
Let us now consider the case  $\epsilon=1$.  The generators take the form 
 \begin{align}
\label{e150}
\hat H&=-\frac{i}{2}\sin\psi-\frac 1 2 (1+\cos\psi)+i(1+\cos\psi)\frac{d}{d\psi},\\
\label{e150a}
\hat K&=\frac{i}{2}\sin\psi-\frac 1 2 (1-\cos\psi)+i(1-\cos\psi)\frac{d}{d\psi},\\
\label{e150b}
\hat D&=\frac{i}{2}\cos\psi-\frac 1 2 \sin\psi+i\sin\psi\frac{d}{d\psi}.
\end{align}
Thus,
\be{151}
f_E(\psi)=\frac{1}{\sqrt{2\pi}}\textrm{sgn}(\pi-\psi)e^{-iE\tan(\frac\psi 2)}e^{-i\frac{\psi}{2}}
(1+\cos\psi)^{-\frac 12}.
\ee
Computing the coordinate representation of the energy eigenfunctions we find 
 \be{152}
\tilde f _E(x)=\left\{
\begin{array}{c}
-i\textrm{sgn}(E)|x|^\frac 12 J_0(\sqrt{2|E|}|x|), \quad E/x>0;\\
0, \quad E/x<0.
\end{array}
\right.
\ee
The reducibility of representation is now clearly seen. The subspace of square  integrable functions supported on positive semiaxis span  the representation  corresponding to  the positive part of the spectrum  of $\hat H$ and $\hat K$.  In fact, this  subspace is obviously invariant under  the global action of $\hat K$ (multiplication by the $x$-dependent phase factor) and $\hat D$  (scaling of independent   variable and  multiplication by $x$-independent factor); the invariance  under  the action  of $\hat H$  follows from  the form  of eigenfunctions. The subspace  of functions 
 supported on negative  semiaxis  carries the representation  corresponding to negative  eigenvalues  of $\hat H$ and $\hat K$. 
Let us note that in accordance with general theory of $SL(2,\mR)$  representations, the wave  functions (\ref{e152})  can be viewed as $m\rightarrow 0$ limit of the eigenfunctions spanning the representations  belonging to the discrete series (cf. eq. (\ref{e57})); the limiting representations are sometimes called the mock representations.
\section{Representations of the universal  covering}
It follows from the previous analysis  that the positive values of coupling constant are quantized. This restriction on the Casimir spectrum follows from the topology of $SL(2,\mR)$ group.  For 
\be{153}
g=\left(
\begin{array}{cc}
a&b\\
c&d
\end{array}
\right),
\ee  
the defining  condition $ad-bc=1$ can be rewritten as 
\be{154}
x_1^2+x_2^2-x_3^2-x_4^2=1,
\ee
with
\be{155}
x_1=\frac 12(a+d),\quad x_2=\frac 12(b-c), \quad x_3=\frac 12(a-d), \quad x_4=\frac 12 (b+c).
\ee
Therefore, the group manifold is  a hyperboloid. It contains the unshrinkable circle
\be{156}
x_1^2+x_2^2=1,\quad x_3=0, \quad x_4=0,
\ee  
which  corresponds to the compact subgroup  generated by $(H+K)$ (cf. eq.  (\ref{e86}))
\be{157}
g(\theta)=\left(
\begin{array}{cc}
\cos\theta &\sin\theta\\
-\sin\theta &\cos\theta
\end{array}
\right).
\ee  
It is not difficult to see that the homotopy group  of $SL(2,\mR)$ is $\mathbf{Z}$ which is  the  homotopy group of a circle.  Eq.  (\ref{e61})  implies for $g$  given by eq.  (\ref{e157}) 
\be{158}
 \left(\tilde D_m^+(g(\theta))f\right)(w)=e^{-i(m+1)\theta} f (e^{-2i\theta}w),
\ee
which  explains the quantization of $m$. On the contrary, eq.  (\ref{e78}) describe the continuous serie and   does not impose any  restriction on $\rho$  following from periodicity  in $\theta$. The same concerns the supplementary serie. 
\par
Admitting the representations  of the universal covering $\widetilde{SL(2,\mR)}$  one can  relax the quantization condition  for the coupling constant in the  "discrete" series.  In the case of continuous series  the parameter $\epsilon$  also ceases  to be discrete. The most interesting case of appearance  of the representations of the universal covering $\widetilde{SL(2,\mR)}$  is probably the case of free theory. Putting $g=0$ in eq. (\ref{e1}) we find 
\be{159}
H=-\frac 12\frac{d^2}{dx^2},\qquad D=\frac{ix}{2}\frac{d}{dx}+\frac i 4,\qquad K=\frac{1}{2}x^2.
\ee
The above operators act in $L^2(\mR)$. Both $\hat H$ and $\hat K$  are positive so only discrete  series  enter the game. It  follows  from eq. (\ref{e58}) (or (\ref{e81})) that $m=\pm\frac 12$ (for the universal covering of $SL(2,\mR)$ and the counterpart of  the discrete series $m>-1$ ). Repeating the reasoning  presented in Section 3 we find the coordinate representation  for $m=\frac 12$ 
\be{160}
\tilde f_E(x)=i^{-\frac 32}\sqrt x J_{\frac 12}(\sqrt{2E}x)=i^{-\frac 32}\sqrt{\frac {2}{\pi\sqrt{2E}}}\sin\sqrt{2E} x, \quad x>0.
\ee 
Putting $m=-1/2$ and choosing the parametrization of the eigenvalues of $\hat K$ as 
\be{161}
\kappa=\frac{x^2}{2}, \quad x<0,
\ee
one easily finds
\be{162}
\tilde f_E(x)=i^{-\frac 12}\sqrt{| x|} J_{-\frac 12}(|x|\sqrt{2E})=i^{-\frac 12}\sqrt{\frac {2}{\pi\sqrt{2E}}}\cos \sqrt{2E} x, \quad x<0.
\ee  
So we are dealing with the representation acting in $L^2(\mR)$  which is reducible as a sum of representations corresponding  to $m=\frac 12$ and $m=-\frac 12$ acting in the subspaces of functions having their supports in right and left semiaxis, respectively.  Now, one can  proceed as follows.  The  functions (\ref{e160}) can be defined on the whole real  axis  by antisymmetry  while  the functions (\ref{e162})  by symmetry. As a result the $m=\frac 12$  and $m=-\frac 12$  representations can be  described as acting in the subspaces  of $L^2(\mR)$  of odd or even  functions, respectively. Our representation defined in $L^2(\mR)$, is obtained by decomposing any element $f\in L^2(\mR)$ into  even and odd parts, $f(x)=\frac 12((f(x)+f(-x))+ \frac 12 (f(x)-f(-x))$ (cf. the results obtained in Ref. \cite{b44b}).
As in the previous cases  one can start with the classical phase space. In the present case it  is a (say, forward) light cone $(\xi_0)^2-(\xi_1)^2-(\xi_2)^2=0$. It is parametrized as follows (cf. eqs.  (\ref{e28})  and (\ref{e29}))   
\begin{equation}
\label{e163}
\begin{split}
\xi_0&=\frac 14 (x^2+p^2),\\
\xi_1&=\frac 14 (x^2-p^2),\\
\xi_2&=-\frac 1 2 xp,
\end{split}
\end{equation}
with $x$ and $p$ being the  Darboux variables. On the quantum level 
\begin{align}
\label{e164}
\xi_0&=\frac 14 (x^2-\frac{d^2}{x^2}),\\
\xi_1&=\frac 14 (x^2+\frac{d^2}{dx^2}),\\
\xi_2&=\frac i 2 x\frac{d}{dx}+\frac i4.
\end{align}
Then one easily checks that
\be{165}
(\hat \xi_0)^2-(\hat \xi_1)^2-(\hat \xi_2)^2=-\frac {3}{16},
\ee
which  perfectly  agrees  with  eq.  (\ref{e79}). Let  us  note that, with  the line $\xi_0=-\xi_1$, $\xi_2=0$ deleted, the cone  can be mapped onto the half-plane $x>0$ ($x<0$), $p\in \mR$:
\be{166}
x=\pm\sqrt{2(\xi_0+\xi_1)}, \quad p=\frac{\mp2\xi_2}{\sqrt{2(\xi_0+\xi_1)}}.
\ee 
However, if we consider the whole $x$-$p$ plane, we are dealing with a double covering  of the cone (cf. eq. (\ref{e31})). This is a classical  counterpart of the construction of quantum theory described above. 
\par 
The full description of quantum systems related to the representations  of the universal covering of $SL(2,\mR)$ will be given elsewhere. 
\section{Discussion}
We have described the quantum version of  the conformal mechanics of one degree of freedom  from the point of view  of the unitary  representations  of the $SL(2,\mR)$ group -- conformal group in ($1+0$)-dimension.   Let us emphasize  the main points of the discussion. 
\par According to the common wisdom the conformal theory  is well-defined in the repelling case, i.e.,  for positive coupling constant $g^2>0$. The Hamiltonian can be then consistently  defined  as a  self-adjoint  operator  generating unitary dynamics.  On the contrary,  the attractive case, $g ^2<0$, is  believed to be  plagued by the "falling on the center" phenomenon.  However, even in this case one can define the self-adjoint  operator  starting from  the formal differential  expression (\ref{e1}). The point is that  the resulting energy spectrum breaks  the conformal  symmetry. One obtains a dynamical system with the conformal  symmetry broken "quantum mechanically" (i.e., we are faced with kind of quantum anomalies). The source of the trouble may be  traced  back to the  classical case.  In the attractive case we are dealing with a kind of pathology: every trajectory  hits  in finite time  (positive or negative) the boundary  $x=0$ of the phase space.
 This conclusion  is, however, misleading. 
 If we define an elementary  conformal  invariant system as the one  described by  the phase space  on which  the conformal  group  acts transitively we can classify such   systems  using the orbits method. It appears that  the phase space for the attractive case   is isomorphic to  one-sheeted hyperboloid which has  nontrivial topology. The  conformal  invariant dynamics  is perfectly  regular,  with no singularities. The apparent  singularity  appears  due to the fact  that  the phase  space, being   topologically nontrivial, cannot be covered  by one map. When proper covering  of phase  space  is constructed the  dynamics  becomes smooth. Once this  fact is properly recognized both classical and quantum dynamics can be quite easily constructed. In the quantum case  the starting point is the choice of  a unitary representation of $SL(2,\mR)$; in this way  the exact conformal symmetry  is built into the theory  from very  beginning. It is then easy  to find  the generators  as well-defined self-adjoint operators. It remains to define the coordinate representation. To this end we invoke  eqs. (\ref{e1b}), (\ref{e16}),  (\ref{e21}) and (\ref{e24}) to define  the $x$ coordinate in terms of the spectrum of $\hat K$. The picture  is closed by showing  that  the canonical  (geometric) quantization of classical theory defined on the relevant coadjoint orbits yields  the quantum picture we have started with.        

\par
{\bf Acknowledgments}
The  author is grateful to  Cezary and Joanna Gonera, Piotr Kosi\'nski and Pawe\l \, Ma\'slanka for their encouragement and stimulating discussions without which  the paper certainly  could not exist.  
The research was supported by the grant of National Science Center number DEC-2013/09/B/ST2/02205.

\end{document}